\documentclass[a4paper,onecolumn,twoside,fleqn,10pt]{article}
\usepackage{amsmath}
\usepackage{authblk}
\usepackage[svgnames]{xcolor}
\definecolor{blue}{RGB}{0,0,255}
\usepackage{caption}
\usepackage[font={color=blue},figurename=Fig.,labelfont={it}]{caption}
\usepackage{bm}
\usepackage{subcaption}
\usepackage[varg]{txfonts}
\usepackage{graphicx}
\usepackage{pdfpages}
\usepackage{wasysym}
\usepackage{rotating}
\usepackage{hyperref}
\usepackage{indentfirst}
\usepackage{float}
\usepackage{lipsum} 
\usepackage{setspace} 
\usepackage{indentfirst}
\usepackage[english]{babel}
\usepackage[switch,columnwise]{lineno}
\usepackage[left=2cm,right=2cm,top=2cm,bottom=2cm]{geometry}
\usepackage{cite}
\usepackage{enumitem}   

\hypersetup{
    colorlinks=true,                          
    linkcolor=blue, 
    citecolor=blue, 
    urlcolor=blue,
    linktoc=all
}

\newcommand{\blue}[1]{\textcolor{blue}{#1}}

\newcommand{\etc}{\textit{etc }\ldots }
\newcommand{\eg}{\textit{eg. }}
\newcommand{\cf}{\textit{cf. }}
\newcommand{\ie}{\textit{ie. }}

\newcommand{\degree}{$^{\circ}$}

\newcommand{\dAlembert}{\blue{d'Alembert} }
\newcommand{\Lagrange}{\blue{Lagrange} }
\newcommand{\Laplace}{\blue{Laplace} }
\newcommand{\LeMouel}{\blue{Le Mouël et al.} }

\newcommand{\Milankovic}{\blue{Milankovi\'c} }
\newcommand{\Laskar}{\blue{Laskar et al.} }

\newcommand{\Lopes}{\blue{Lopes et al.} }
\newcommand{\Scafetta}{\blue{Scafetta et al.} }
\newcommand{\Courtillot}{\blue{Courtillot et al.} }
\newcommand{\Gupta}{\blue{Gupta et al.} }
\newcommand{\Kristoufek}{\blue{Kristoufek et al.} }
\newcommand{\Dai}{\blue{Dai et al.} }
\newcommand{\Connolly}{\blue{Connolly et al.} }
\newcommand{\Ormaza}{\blue{Ormaza-González et al.} }
\newcommand{\Li}{\blue{Li et al.} }
\newcommand{\ScafettaBianchini}{\blue{Scafetta and Bianchini} }

\newcommand{\Lambeck}{\blue{Lambeck} }
\newcommand{\Stephenson}{\blue{Stephenson and Morrison} }
\newcommand{\Gross}{\blue{Gross} }

\newcommand{\Lemmerling}{\blue{Lemmerling and Van Huffel}}
\newcommand{\Golub}{\blue{Golub and Reinsch}}
\newcommand{\Markowitz}{\blue{Markowitz}}
\newcommand{\Stoyko}{\blue{Stoyko}}
\newcommand{\Chandler}{\blue{Chandler}}

\usepackage{fancyhdr}
\begin{document}
\title{On variations of global mean surface temperature: When Laplace meets Milankovi\'c}

\author[1]{Courtillot Vincent}
\author[1]{Lopes Fernando}
\author[2]{Gibert Dominique}
\author[3]{Boul\'e Jean-Baptiste}
\author[1]{Le Mouël Jean-Louis}

\affil[1]{Universit\'e Paris Cité, CNRS UMR 7154, F-75005 Paris, France}
\affil[2]{LGL-TPE - Laboratoire de Géologie de Lyon - Terre, Planètes, Environnement, Lyon, France}
\affil[3]{CNRS UMR7196, INSERM U1154, Museum National d'Histoire Naturelle, Paris, F-75005,  France}

\date{\today}

\maketitle
	\abstract{In his mathematical theory of climate, Milankovic finds a link between the heat received by the Earth surface per unit time as a function of the solar ephemerids and derives a model of climate changes at periods longer than a few (tens of) thousand years and more. In this paper, we investigate the potential connections of global temperature and Earth rotation at much shorter periods, in the complementary range of one to a few hundred years. For temperature, we select the HadCrut05 data set from 1850 to the Present. For Earth rotation, defined by pole coordinates and length of day, we use the IERS data sets from 1962 to the Present and from 1832 to the Present (annual sampling). Using iterative Singular Spectrum Analysis (iSSA), we extract the trend and quasi-periodic components of these time series. The trends of lod and temperature are anti-correlated, the former decreasing by 1.8 ms when the latter increases by 1.3°C. The first quasi-periodic components (period ~80-90 years) are expressions of the Gleissberg cycle and are identical (at the level of uncertainty of the data). Taken together, the trend and Gleissberg components allow one to reconstruct 87\% of the variance of the data for lod and 48\% for temperature. The next four iSSA components, with periods $\sim$40, 22, 15 and 9 years, match the now well-known list of commensurable periods of the Jovian planets. There is an ongoing debate on the origin of the forcing of these components (astronomical vs Earth-bound). The Lagrange and Laplace theories imply that the derivative of pole motion should be identical to lod variations: this strong check is passed by the trend + Gleissberg reconstructions. The annual oscillations of pole motion and lod are linked to annual variations in Sun-Earth distance, in agreement with an astronomical, but not a climatic origin. The results obtained in this paper for the observed temperature/rotation couple add to the growing list of evidence of solar and planetary forcings of gravitational nature on a number of geophysical processes (including sea-level, sea-level pressure, sea-ice extent, oceanic climate indices). The components found in this paper can be considered as a shorter-period extension of the Milankovic cycles: the excellent phase coincidence between annual cycles of temperature and lod implies Lagrangian forcing of temperature at this period, that is the very idea of Milankovic.}

\section{\label{sec01} Introduction}
	Despite difficulties in defining and measuring the Earth’s “mean global temperature”, there is a broad consensus on its variations since the beginning of the industrial era in the mid-1800s. 
\begin{figure}[H]
    \centering
    \includegraphics[width=1\columnwidth]{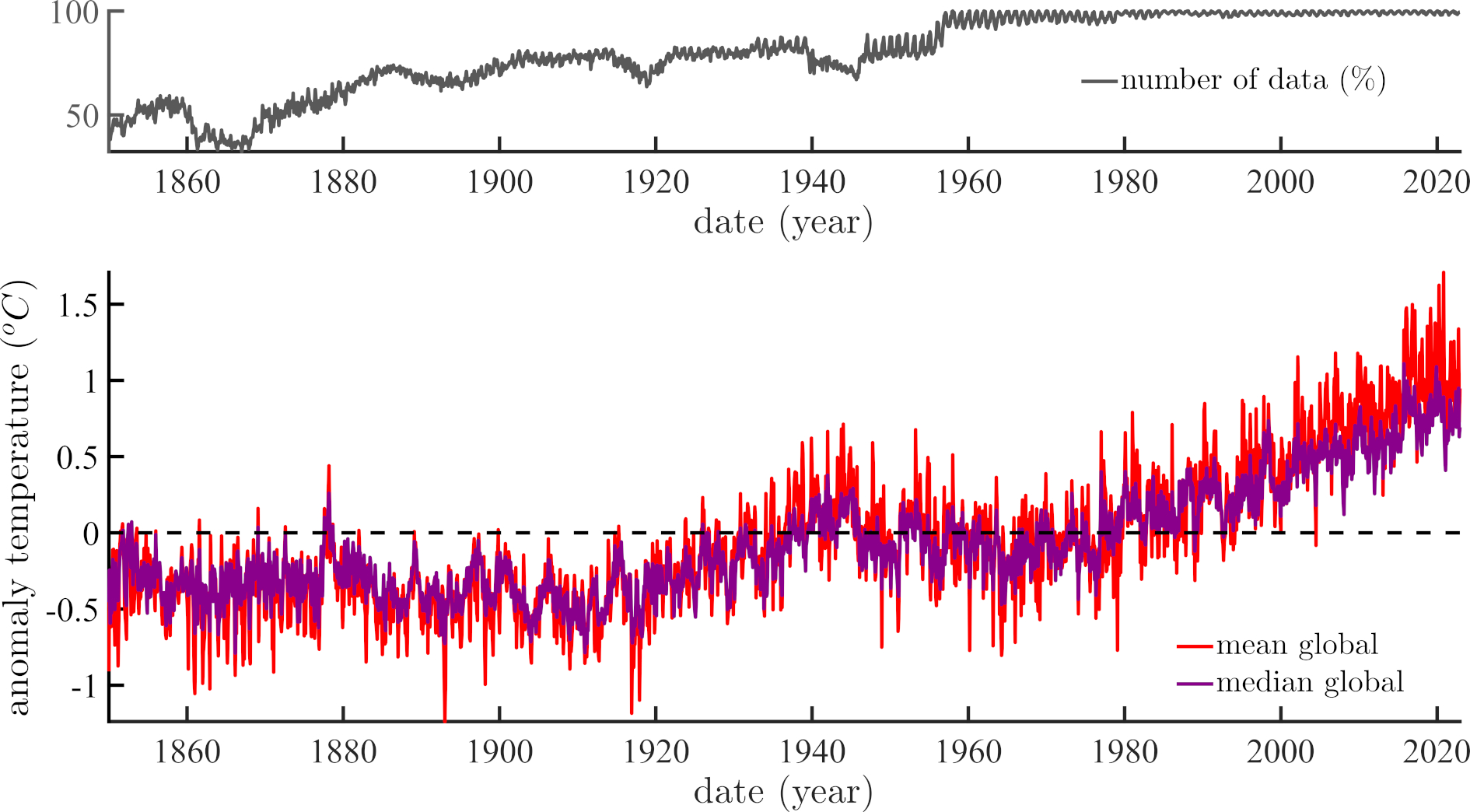}	
    \caption{(top), percent surface coverage of the mean temperature data since 1850 (grey) ; (bottom),  mean (red) and median (purple) temperature anomalies from data set HadCrut 05.}
	\label{Fig:01}
\end{figure}
	
	 We select as a representative data set the HadCrut\footnote{https://crudata.uea.ac.uk/cru/data/temperature/} file (\cf \cite{Morice2021,Osborn2021}) version 05, the one used in the latest IPCC AR6 report (\cf \cite{Masson2021}), shown here in Figure \ref{Fig:01} (bottom). General interest focuses on the warming segments (1920-1940 and 1980-2000 or 2020). One should recall that the data coverage is far from complete before 1960 (Figure \ref{Fig:01} top). This is one of the many reasons for discrepancies between observed and modelled temperatures (\eg \cite{Thorne2007,Douglas2008,Santer2008}). The temperature anomalies are an indicator of temperature variations.

	In his mathematical theory of climates, \Milankovic (\cite{Milankovic1920}, page 15, equation 20) proposes a link between the amount of heat $\dfrac{d\mathcal{W}}{dt}$ received by the Earth’s surface by unit time, at a location with latitude $\varphi$ and longitude $\psi$ and as a function of the Sun’s declination $\delta$ and and hour angle $\omega$,
\begin{equation*}
	\dfrac{d\mathcal{W}}{dt} = \dfrac{\mathcal{I}_0}{\rho ^2} [\textrm{sin}\ \varphi \ \textrm{sin}\ \delta + \textrm{cos}\ \varphi \ \textrm{cos}\ \delta \ \textrm{cos}\ (\omega + \psi)]
\end{equation*}

	where $I_0$ is the solar constant and $\rho$ the Sun-Earth distance. \Lopes (\cite{Lopes2022a}) have extended the use of this equation to shorter periods in the range from less than 1 year up to 300 years. As first shown by \Laplace (\cite{Laplace1799}), the Sun’s ephemerids ($\omega$,$\delta$) are connected directly to variations in pole rotation, that is to the coordinates of the pole ($m_1$,$m_2$) and to the length of the day (\textbf{lod}) (\eg \cite{Lambeck2005,Lopes2022b,Lopes2022c}). It has been known for decades (\eg \cite{Morth1979}), with renewed interest in recent years (\eg \cite{Bank2022,Lopes2022b,Scafetta2022}), that many geophysical and astrophysical time series involve the same sets of pseudo-periodical components, all of them connected to commensurable periods of the planets. The very same pseudo-periodic and periodic components, found in many geophysical phenomena (\eg \cite{Lyons1899,Bartels1932,Morth1979,LeMouel2019b,LeMouel2019c,Dumont2020,Zaccagnino2020,Dumont2021,Lopes2021b,Dumont2022,Petrosino2022,Zaccagnino2022,LeMouel2023}) are also found in proxies of solar activity (\eg \cite{Stefani2016,Stefani2019,LeMouel2020b,Scafetta2020,Courtillot2021}). And for the specific problem of variations in climate-sensitive indices there is an already long list of examples: \Courtillot (\cite{Courtillot2013}), \Scafetta (\cite{Scafetta2013a,Scafetta2013b,Scafetta2014}), \Gupta (\cite{Gupta2015}), \Kristoufek (\cite{Kristoufek2017}), \Dai (\cite{Dai2019}), \LeMouel (\cite{LeMouel2019a,LeMouel2020a}), \Connolly (\cite{Connolly2021}), \LeMouel (\cite{LeMouel2021}), \Lopes (\cite{Lopes2021a}), \Courtillot (\cite{Courtillot2022a,Courtillot2022b}), \Li (\cite{Li2022}), \Lopes (\cite{Lopes2022b}), \Ormaza (\cite{Ormaza2022}), \Lopes (\cite{Lopes2023a}) and \ScafettaBianchini (\cite{Scafetta2023}). The main remaining problem is to find whether these correlations (sharing the same sets of periodic and pseudo-periodic components) imply causal physical mechanisms. One can provide a partial answer to this important question, at least in the field of geophysics. In most other scientific disciplines, particularly those involving data mining, knowledge discovery, or pattern recognition, determining whether certain patterns are related involves the comparison of their Fourier spectra, of their wavelet coefficients, \etc (\eg \cite{Keogh2001,Lin2003}). That is precisely what is proposed here.

	\Milankovic (\cite{Milankovic1920}) imagined that the mechanism that could explain the distribution of climates on Earth had to be geometrical. Recall that in ancient greek $\kappa\lambda\iota\mu\alpha$ (pronounced clima) meant “inclination of Earth towards the pole starting at the equator”. We have known since \Lagrange (\cite{Lagrange1788}) and \Laplace (\cite{Laplace1799}) that planets behave as tops and that variations in their inclination axis and rotation velocity (or length of day) are linked. \Milankovic (\cite{Milankovic1920}), and more recently \Laskar (\cite{Laskar1993}) relate insolation and variation in obliquity of the Earth’s rotation axis, that itself is related to this rotation. In this paper we pursue along those lines, with the temperature/rotation couple in mind.
	
	Following the present introduction (section \ref{sec01}), section \ref{sec02} is devoted to a discussion of the data on Earth rotation and the length of day. Periodic and semi-periodic components of temperature and lod are extracted by singular spectral analysis (iSSA) and compared in section \ref{sec03}. The first two components are presented in some detail, followed by a sub-section in which the derivative of polar coordinates is shown to be a proxy of lod, leading to some interesting results. The discussion and conclusions form section \ref{sec04}.

\section{\label{sec02} Earth rotation and the length of day}
	The Earth’s rotation involves three co-ordinates, the two coordinates of the rotation pole and the rotation velocity or the related length of day (\textbf{lod}). We refer the interested reader to the reference book by \Lambeck (\cite{Lambeck2005}) (chapter 5.2, page 73). There is an ongoing debate between the original treatment of what is now called the Liouville-Euler system of equations by pioneers \Lagrange (\cite{Lagrange1788}) and \Laplace (\cite{Laplace1799}) and the way they are understood by a number of specialists in geodesy (\Lambeck \cite{Lambeck2005}, chapter 3). Our contribution to this debate is in \Lopes (\cite{Lopes2022c}). As far as lod is concerned, the two approaches differ on the origin of the forcing, involving mainly astronomical sources for some (\cite{LeMouel2019b,Lopes2022b}) or earth-bound sources for others (\cf \Lambeck \cite{Lambeck2005}, page 34). 
		
\begin{figure}[H]
    \centering
    \includegraphics[width=0.8\columnwidth]{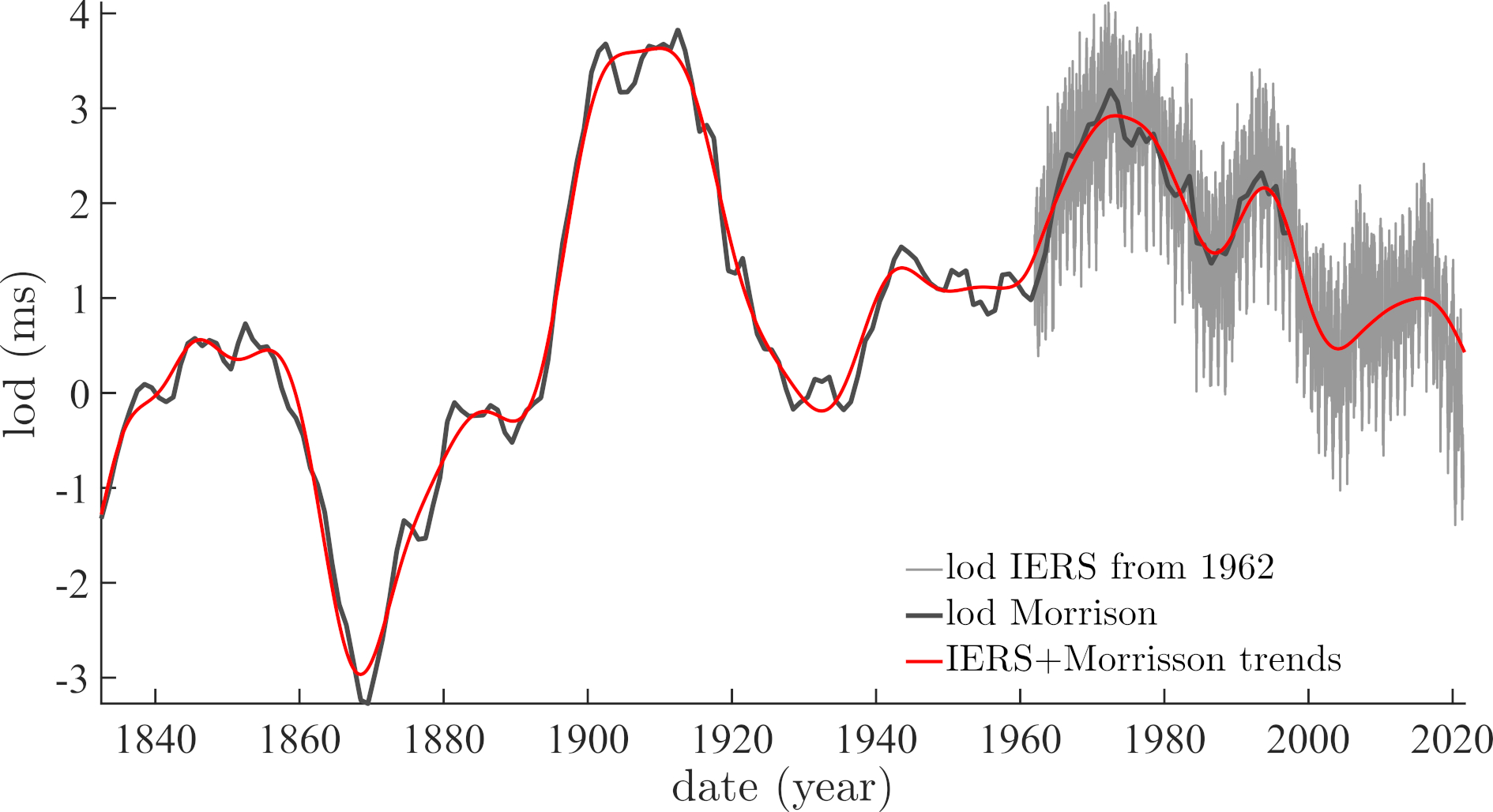}	
    \caption{Length of day. Black curve: annual means from \Stephenson (\cite{Stephenson1984}) and \Gross (\cite{Gross2001}). Grey curve: data from \textbf{IERS} satellites, since 1962. Red curve: cubic spline interpolation as explained in \Lopes (\cite{Lopes2022b}).}
	\label{Fig:03}
\end{figure}

	The \textbf{lod} has been measured by satellites with great precision daily since 1962. The data is accessed through the site of the International Earth Rotation and Reference Systems service (\textbf{IERS}\footnote{https://www.iers.org/IERS/EN/DataProducts/EarthOrientationData/eop.html}). There is also a time series going back to 1832, with annual sampling (\cite{Stephenson1984,Gross2001}). We have combined the two series and generated a suite of annual means (red curve in Figure \ref{Fig:03}), as we had already done in \Lopes (\cite{Lopes2022b}).	
	
\begin{figure}[H]
    \centering
    \includegraphics[width=0.8\columnwidth]{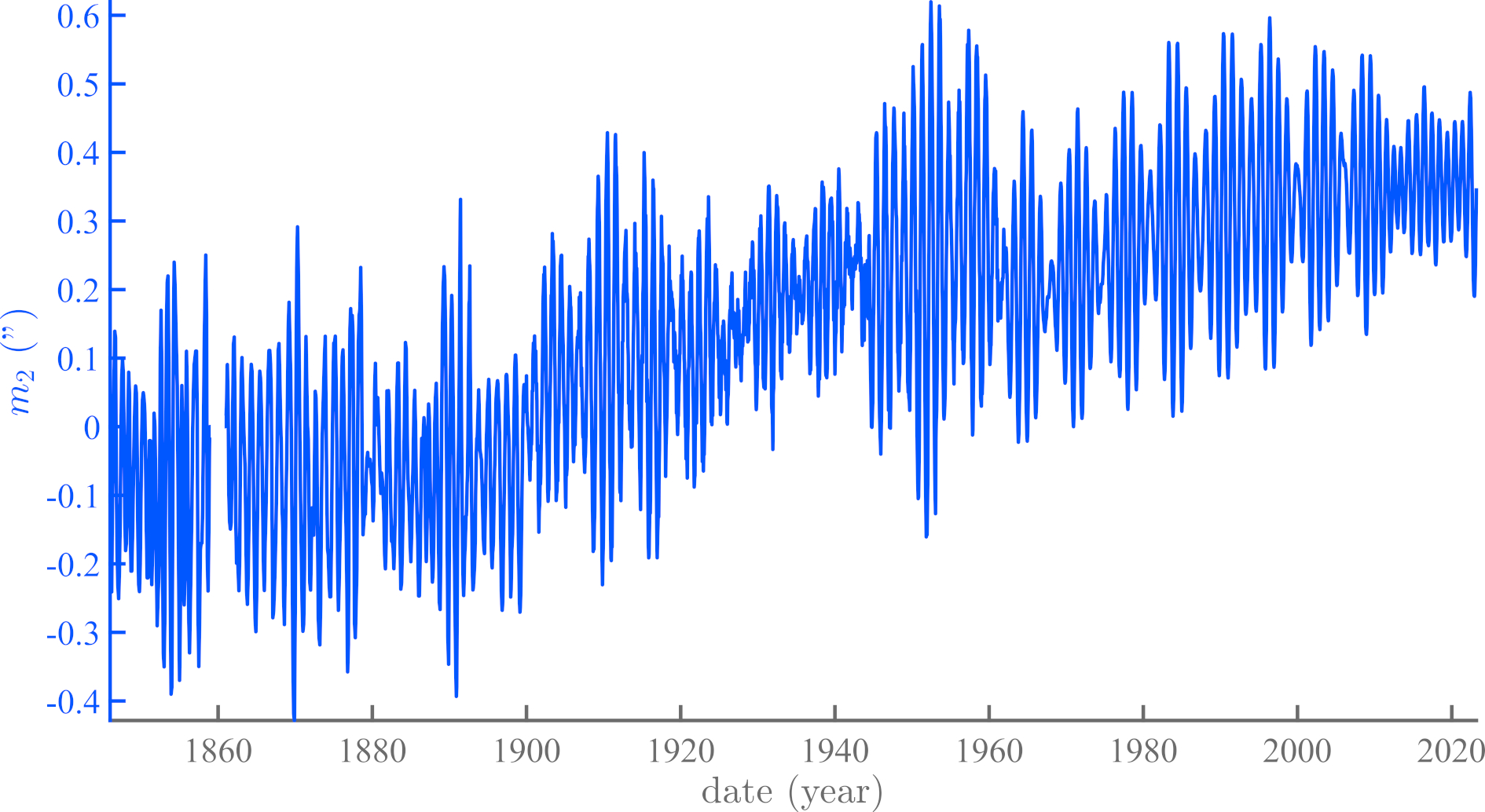}	
    \caption{Evolution of the $m_2$ component of polar motion since 1846.}
	\label{Fig:07}
\end{figure}

		On Figure \ref{Fig:07}, we present the time evolution of the $m_2$ component of the polar motion since 1846; these values are also provided by the IERS. Pole motion is defined by the couple of components ($m_1$ , $m_2$) in the plane ($O$, $m_1$, $m_2$) which is tangent to the North Pole (or in the equatorial plane). O represents a conventional origin, the axis $O$, $m_1$ lies along the Greenwich meridian, and the axis $O$, $m_2$ lies along the 90\degree E meridian (see \cite{Lopes2017,Lopes2022c} for more details). 
		
\section{\label{sec03} Extraction and comparison of pseudo-periodic components} 		
	We have applied \textit{iterative Singular Spectrum Analysis} (\textbf{iSSA}, \eg \cite{Golyandina2013}) in order to extract the trend (first component) and the first periodic and pseudoperiodic components (next sub-section) from the time series of global temperature and length of day. \textbf{SSA} uses the mathematical properties of descending order diagonal matrices (Hankel, Toeplitz matrices; see \Lemmerling \cite{Lemmerling2001}) and their orthogonalization by singular value decomposition (\textbf{SVD}, \Golub  \cite{Golub1971}).

\subsection{The first two components}
	A quick look at the relative amounts of information carried by the monthly mean curves of temperature HadCrut5 (Figure \ref{Fig:01}) versus \textit{lod} (Figure \ref{Fig:03}) leads one to conclude that the comparison is unlikely to be worthwhile beyond three components. Let us look at the first two.
\begin{figure}[H]
    \centering
    \includegraphics[width=0.8\columnwidth]{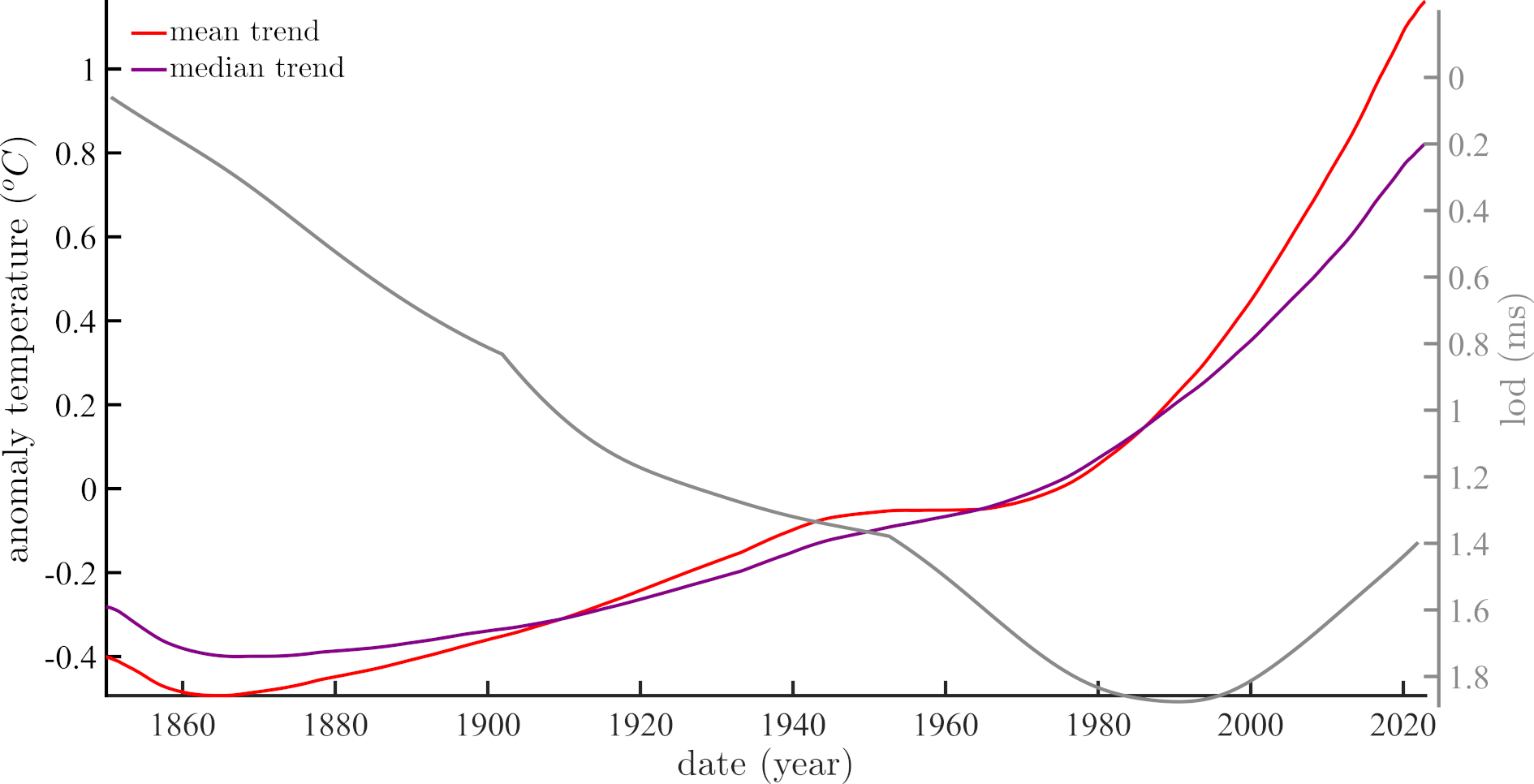}	
    \caption{Trends (first component) extracted with \textbf{iSSA} for respectively mean (red) and median (purple) temperatures and \textit{lod} (grey).}
	\label{Fig:04}
\end{figure}
\
\begin{figure}[H]
    \centering
    \includegraphics[width=0.8\columnwidth]{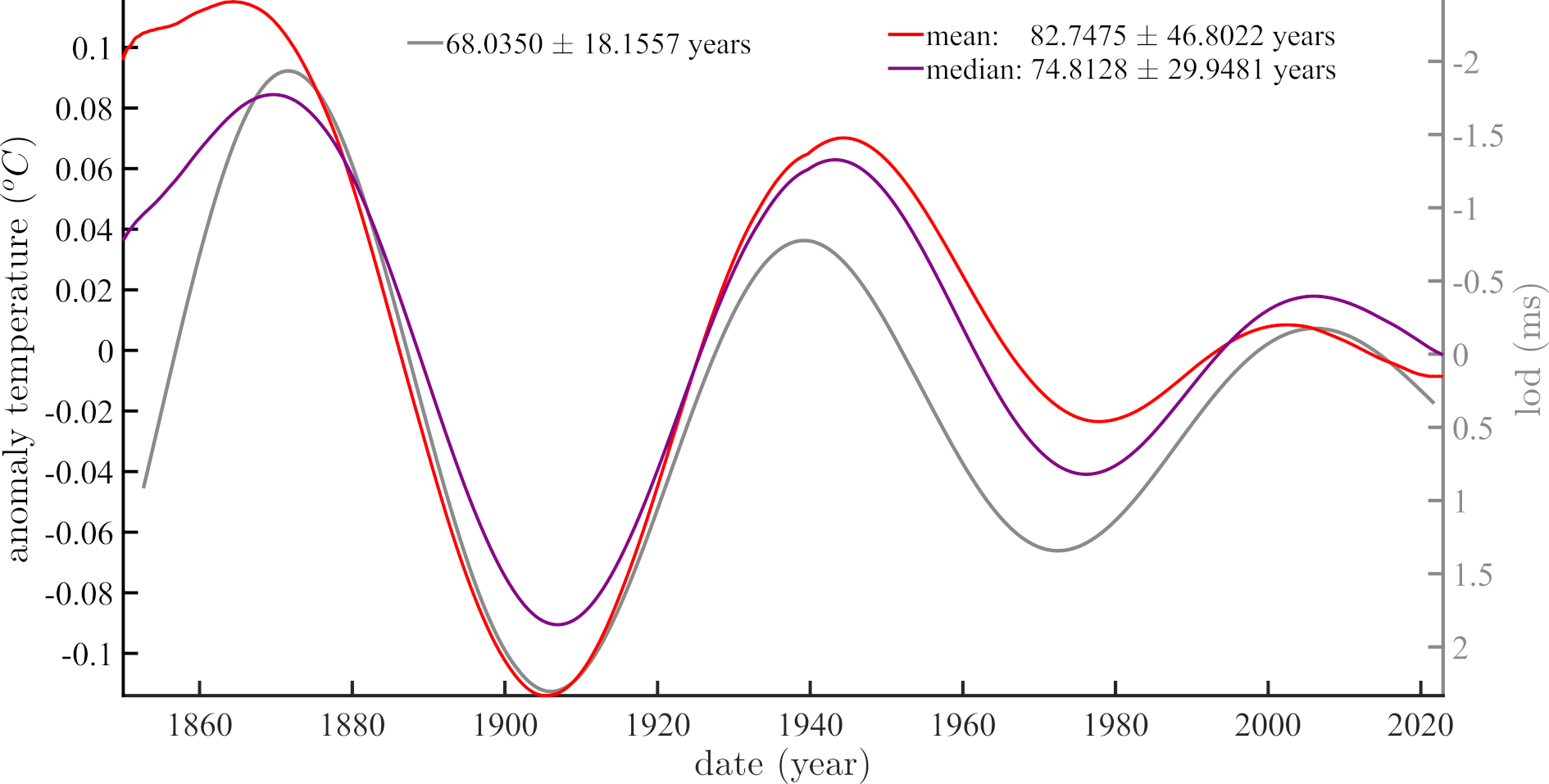}	
    \caption{$\sim$80 to $\sim$90 year \textbf{iSSA} components (Gleissberg cycle) for respectively mean (red) and median (purple) temperatures and \textit{lod} (grey). In this and similar figures the mean period of the \textbf{iSSA} component (in years) is indicated next to the relevant curve color, together with an estimate of uncertainty equal to the width of the component at mid-height.}
	\label{Fig:05}
\end{figure}

\begin{figure}[H]
    \centering
    \includegraphics[width=1\columnwidth]{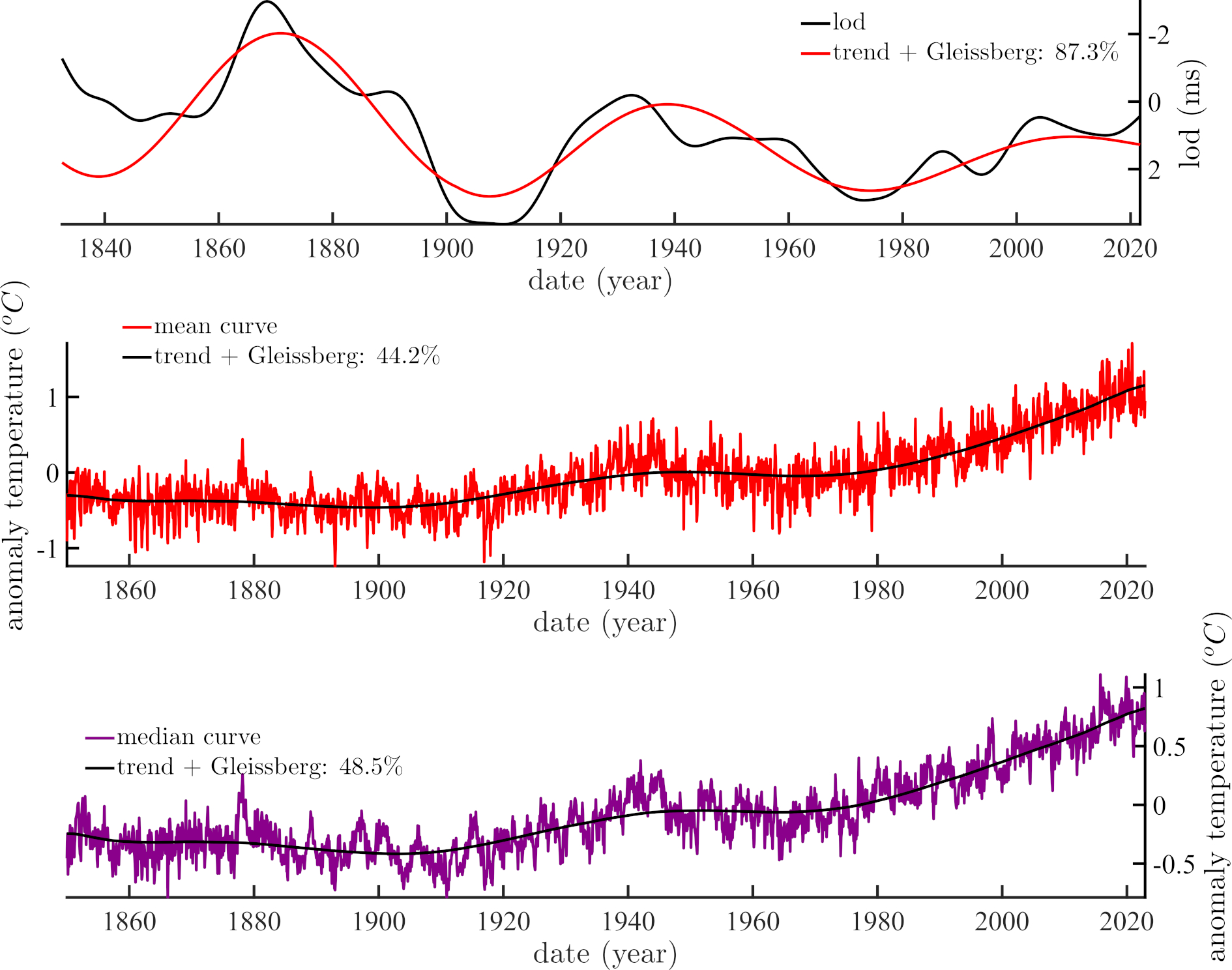}	
    \caption{Reconstruction and calculated variances for the simple addition of trend and Gleissberg cycle. From top to bottom: lod of \Stephenson (\cite{Stephenson1984}), mean HadCrut 05  temperature and median HadCrut5 temperature.}
	\label{Fig:06}
\end{figure}

	Figures \ref{Fig:04} and \ref{Fig:05} display the first two \textbf{iSSA} components (trend and $\sim$80 to $\sim$90 year Gleissberg cycle) for the three time series of temperature (mean and median) and \textit{lod}. As usual, we will draw no conclusions from the trends, which can be represented mathematically by either high-order polynomials or pseudo-cycles. One can guess some long term anti-correlation between the two low-degree trends (Figure \ref{Fig:04}), with superimposed weak multi-decadal oscillations. On the other hand, the correlations of what we interpret as expressions of the Gleissberg cycle (Figure \ref{Fig:05}) are truly remarkable, and the main finding of this paper. We show in Figure \ref{Fig:06} the reconstructed curves with only the trend and Gleissberg cycle included; the variance of the data captured by this simple model is respectively 87\% for \textbf{lod}, and 44 and 48\% for the mean and median temperatures respectively.
	
\newpage
	
\subsection{The derivative of $m_2$ (IERS rotation pole coordinate) as a proxy of \textbf{lod}.}	
	As seen in Figure \ref{Fig:05}, the correlation of the Gleissberg cycles of temperature and length of day is remarkable both in phase and amplitude. In \Lopes (\cite{Lopes2022c}) and \LeMouel (\cite{LeMouel2023}), we have shown (based on actual observations) that \textit{lod} and mean polar motion should be linked by a simple derivative operator, as demonstrated in theory by \Lagrange (\cite{Lagrange1788}) and \Laplace (\cite{Laplace1799}). Therefore, we have computed the derivative of $m_2$ (or $m_1$ as well) and extracted its \textbf{iSSA} components as in \Lopes (\cite{Lopes2021a}). Indeed, Figure \ref{Fig:07} confirms that the derivative of  is an excellent proxy of lod variations since 1846.	

\begin{figure}[H]
    \centering
    \includegraphics[width=0.7\columnwidth]{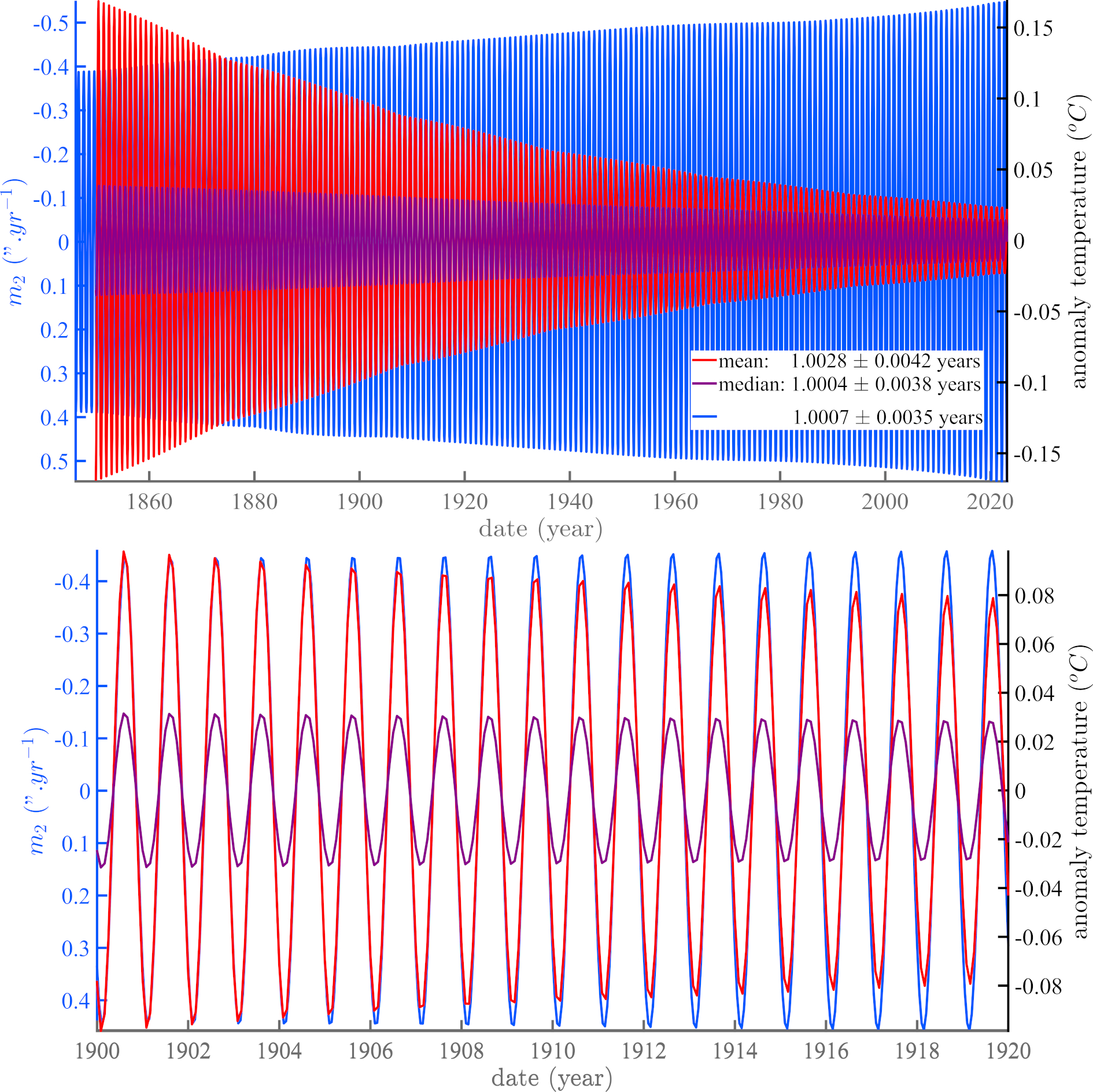}	
    \caption{(top): Forced annual oscillation of the derivative of $m_2$ a proxy of lod (in blue), and of the (resp. mean and median) HadCRUT5 temperatures (resp. purple and red). (bottom): 1900-1920 zoom.}
	\label{Fig:08}
\end{figure}

	The first three components of polar motion are in decreasing order the trend, also called \Markowitz (\cite{Markowitz1968})-\Stoyko (\cite{Stoyko1968}) drift, the \Chandler (\cite{Chandler1891a,Chandler1891b}) free oscillation and the forced annual oscillation (\eg \Lambeck \cite{Lambeck2005} chapter 7; \cite{Courtillot2022a,Lopes2023a}). In comparison, the temperature curve has a trend (discussed in the previous section) and an annual component but no equivalent of the Chandler oscillation. The forced annual oscillations of mean and median temperatures and length of day can be compared in Figure \ref{Fig:08}. The temperature and \textit{lod} components are in phase opposition (note that the positive direction of the $m_2$ ordinate is arbitrary). We have shown in \Lopes (\cite{Lopes2022c}) that the annual oscillations of pole motion and lod were directly linked (as in the Lagrange theory) to annual variations of the Sun-Earth distance. We see here that such remains the case, at least as far as the phases of temperature components are concerned. This is in agreement with an astronomical/gravitational origin of the annual pole motion as understood by \Milankovic (\cite{Milankovic1920}), but on the contrary in contradiction with a climatic origin as preferred by \Lambeck (\cite{Lambeck2005}, chapter 7, page 146) "\textit{The principal seasonal oscillation in the wobble is the annual term which has generally been attributed to a geographical redistribution of mass associated with meteorological causes}".
	
	After having shown the trend and Gleissberg ($\sim$90 yr) components, we briefly discuss the next four \textbf{iSSA} components common to \textit{lod} and HadCrut5 temperatures, in order of decreasing period. They are displayed in Appendix \ref{sec:appendixA} in order to make the paper lighter. These components have periods of  $\sim$40, 22, 15 and 9 years (Figures \ref{Fig:A01} to \ref{Fig:A04}). The components common to temperature and the polar motion, with periods of $\sim$40 and $\sim$9 years, appear to more or less remain in phase throughout the entire observation series. This is not the case for the cycles of $\sim$22 and $\sim$15 years. We observed the same behavior regarding the phases, for the same common pseudo-cycles, when we compared the pole motion to the global volcanic eruption series (see \cite{LeMouel2023b}).

	Table \ref{Tab:02} lists the periods of the periodic (or pseudo-periodic) components common to the three time series, together with their uncertainties.  This list is now well known and misses only a $\sim$30 yr component. A more exhaustive list is given by \LeMouel (\cite{LeMouel2019b}) and \Lopes (\cite{Lopes2021a}) for lod and polar motion, and by \LeMouel (\cite{LeMouel2020a}) for HadCrut5 temperatures.
	
\begin{figure}[H]
    \centering
    \includegraphics[width=0.69\columnwidth]{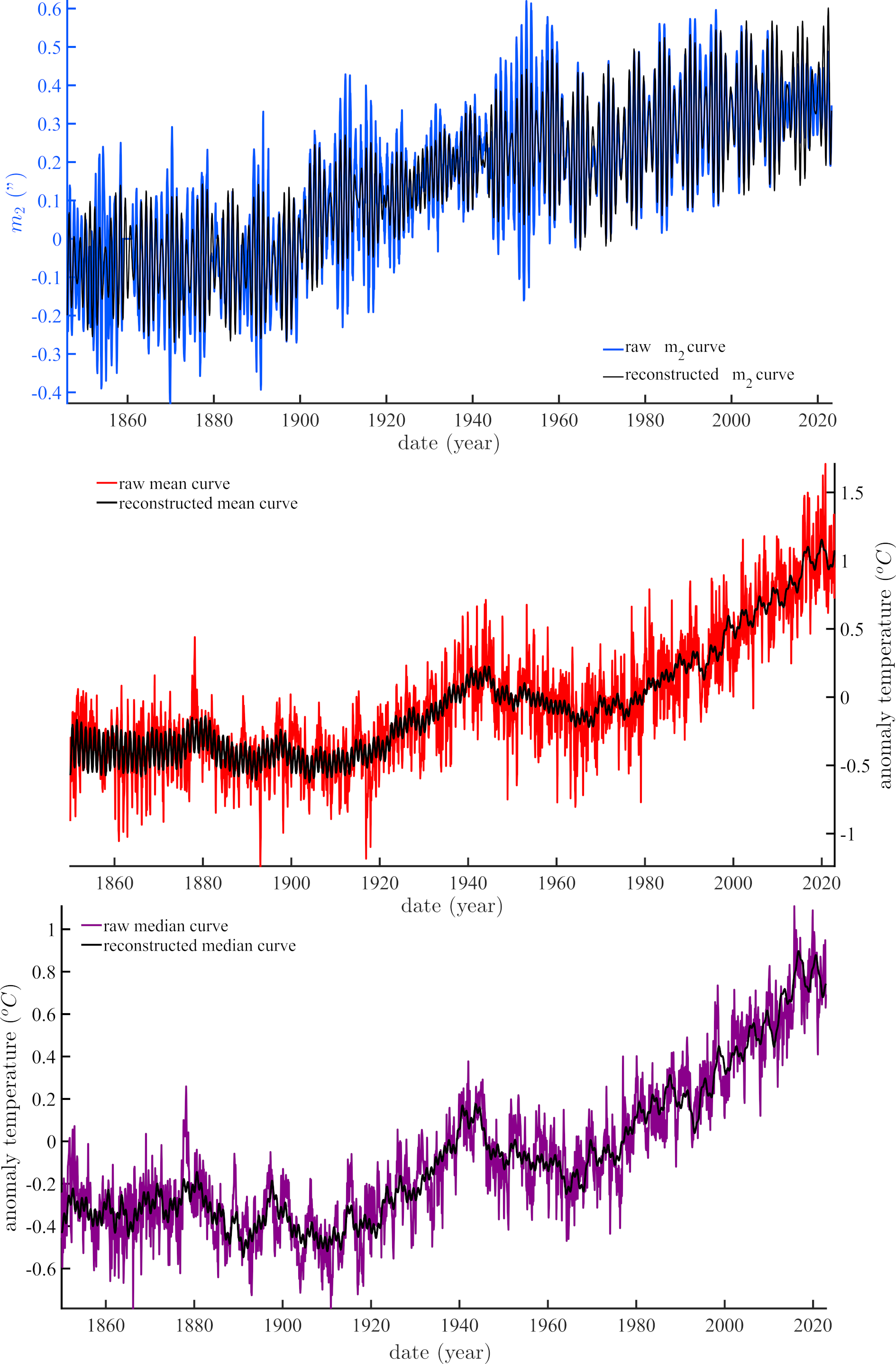}	
    \caption{ Reconstructed (black) over original (color) curves of polar motion (top, blue), mean (red) and median (purple) HadCrut05 temperatures.}
	\label{Fig:09}
\end{figure}

	When one adds the 6 common components, the reconstructed series capture respectively 49\% (mean temperature), 63\% (median temperature), and 69\% ($m_2$ component of polar motion) of the original series variance. The reconstructed curves are shown in black in Figure \ref{Fig:09}, where the trends have been added.

\begin{table}[H]
\centering
\begin{tabular}{ p{2.5cm}|p{2.5cm}| p{2.5cm}}
 \multicolumn{3}{c}{} \\
 \hline
  \hline
  $\dfrac{dm_2}{dt}$  &  mean curve &  mediane curve\\
 \hline
 \ & \ \\
68.03 $\pm$ 18.16 & 82.75 $\pm$ 46.80 & 74.81 $\pm$ 29.95\\
39.01 $\pm$ 5.44  & 37.41 $\pm$  4.95 & 37.66 $\pm$  5.29\\
22.27 $\pm$ 2.14  & 20.77 $\pm$  1.67 & 20.61 $\pm$  1.61\\
14.46 $\pm$ 0.07  & 15.01 $\pm$  0.80 & 14.68 $\pm$  0.85\\
9.10  $\pm$ 0.32  &  9.21 $\pm$  0.30 &  9.12 $\pm$  0.29\\
1.00  $\pm$ 0.01  &  1.00 $\pm$  0.01 &  1.01 $\pm$  0.01\\
 \hline
\end{tabular}   
    \caption{\textbf{iSSA} components (without the trends) of the three time series $\dfrac{dm_2}{dt}$ (a proxy of \textit{lod}), mean HadCrut5 temperature and median HadCrut5 temperatures.}
    \label{Tab:02}
\end{table}

\section{\label{sec04} Discussion and Conclusion}
	This paper is part of a series in which we find strong evidence of solar and planetary forcings of gravitational nature on a number of natural geophysical processes, such as variations of length of day, sea-level, sea-level pressure, extent of sea-ice, the main climatic indices, $\ldots$ In the present paper, we focus on the database of surface temperature anomalies that is put forward in the latest IPCC AR6 report (\cite{Masson2021}). 
	
	We have first recalled that the \Milankovic ’s (\cite{Milankovic1920}) mathematical theory of climates links the rotation and revolution periods of planets to the amount of heat (insolation) they receive (see also \Laskar  \cite{Laskar1993,Laskar2004}; \Lopes \cite{Lopes2022a}). The inclination of the rotation axis and the amount (duration) of insolation force the temperature at any point given by its latitude and longitude, as envisioned by \Laplace (\cite{Laplace1799}).We have recently extended the \Milankovic ’s theory to the range of periods from 0 to 300 years
	
	In order to estimate how much variations of planetary \textit{rotation} could influence (force) surface temperature variations, as represented by the HadCrut5 data set, we have analyzed with \textbf{iSSA} the historical curves of \textit{lod} from \Stephenson (\cite{Stephenson1984}) and \Gross (\cite{Gross2001}).  We have, somewhat conservatively, limited a detailed analysis to the first two components, the trend and the $\sim$90 yr Gleissberg cycle (\cite{Gleissberg1939,LeMouel2017}).  As shown in Figure \ref{Fig:06}, the sum of these two components accounts for $\sim$50\% of the temperature variance. Figure \ref{Fig:04} shows that the trends of \textit{lod} and temperature are anti-correlated, the former decreasing by 1.8 ms as the latter increases by 1.3\degree C. The most remarkable result is displayed in Figure \ref{Fig:05} in which the Gleissberg cycles of length of day and temperature anomaly are seen to be almost perfectly correlated in phase and amplitude (modulation) over more than two cycles (since 1850). Note that given the uncertainties the periods of all three curves can be reasonably assumed to be the same.
	
	In a previous study (\Lopes \cite{Lopes2021a}), we have shown that the two components of polar motion ($m_1$, $m_2$) can be decomposed on a base of pseudo-cycles with the same periods (see Table \ref{Tab:02}). We also know that temperature and \textit{lod} share a number of more minor cycles (\eg \cite{Lopes2017,LeMouel2020a,Lopes2021a,Courtillot2021}). The extracted components with periods of $\sim$90, $\sim$40, $\sim$30, $\sim$22, $\sim$15, $\sim$9 and $\sim$1 yr) are displayed in Figures \ref{Fig:05}, \ref{Fig:A01}, \ref{Fig:A02}, \ref{Fig:A03}, \ref{Fig:A04} and \ref{Fig:08}. Disregarding the trends, these 7 components represent respectively 49\%, 63\% and 69\% of the original signal variance of $m_2$. All  the periods encountered are within uncertainty planetary commensurable periods, that is connected to the four giant (Jovian) planets (\eg \cite{Morth1979,Bank2022}).
	
	The annual oscillation of temperature (and in general of climate and also of pole motion) is generally considered as evidence of a climatic forcing (\eg \cite{Lambeck2005}, chapter 7). But we (\cf \cite{Courtillot2022a,Courtillot2022b,Lopes2022a,LeMouel2023,Lopes2023a}) and others have shown that this forced annual oscillation, be it recorded as variations of sea-ice extent, of sea-level, of \textit{lod}, of polar motion, and even of the geomagnetic field, is clearly and precisely connected to the Earth ephemerdis in full agreement with the theories of \Lagrange (\cite{Lagrange1788}) and \Laplace (\cite{Laplace1799}). The excellent phase coincidence shown in the zoom of Figure \ref{Fig:08} (bottom) between the mean and median annual cycles of temperature and that of length of day (an external parameter as recalled by \Lambeck (\cite{Lambeck2005}, chapter 03) implies a Lagrangian forcing of temperature for this precise period. This is the very idea of \Milankovic (\cite{Milankovic1920}). Thus, one can conclude that reorganization of fluid masses is not the origin of this cycle.
	
\begin{figure}[H]
    \centering
    \includegraphics[width=0.8\columnwidth]{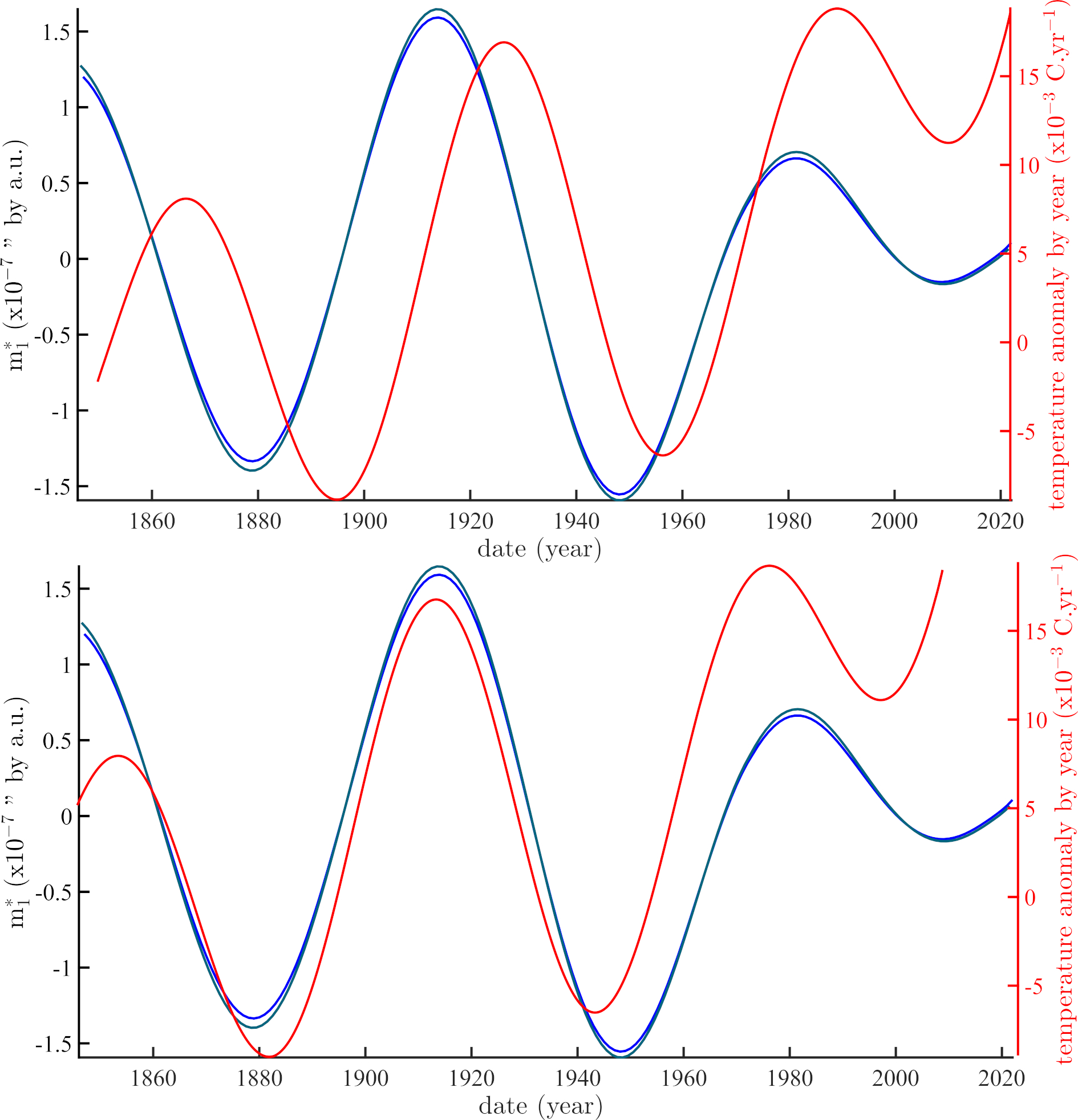}	
    \caption{(from \Lopes (\cite{Lopes2022a}). In red the time derivative of the \textbf{iSSA} trend of temperature anomaly HadCrut5; in blue the inverse distance of terrestrial solstices. Top: the original curves; bottom: the red curve offset by 20 years is brought into phase with the blue curve.}
	\label{Fig:10}
\end{figure}

	We recall that in \Lopes (\cite{Lopes2022a}) we showed that the trend (mean drift) of temperatures is closely connected to the precession of solstices, as predicted by \dAlembert (\cite{dAlembert1749}). Given \Milankovic ’s (\cite{Milankovic1920}) equation 20 (page 15), recalled in section \ref{sec01}, the mean insolation, \ie amount of heat received by Earth in a unit of time, is proportional to the inverse square of distance to the Sun. Thus, in \Lopes (\cite{Lopes2022a}) we obtained Figure \ref{Fig:10}, in which the time derivative of the trend of temperature anomaly HadCrut5 is seen to correlate very well (with a phase difference) with the inverse of distance at the terrestrial solstices.

	These two lines of observation (and reasoning) fully validate the joint use of \Laplace ’s (\cite{Laplace1799}) and \Milankovic ’s (\cite{Milankovic1920}) theories regarding measured surface temperatures. Variations in rotation and revolution lead to short period changes in temperature (and climate in general), whereas variations in the obliquity, precession and eccentricity of the rotation axis lead to much longer periods.
	
	For \Laplace as well as for \Milankovic, and more recently Laskar, who are interested in the longer periods, characteristic of paleo-climate changes, the parameters whose paleo-variations matter most, the important (i.e. the precession of equinoxes and longer) parameters are the position and orientation of the planet’s axis of rotation that governs the amount of heat received by the planet, hence its temperature. But for shorter time scales, as implied by Milankovic’s equation, one is dealing with climate at shorter periods (the ones we list from 1 to 90 years in particular).

	Very unexpectedly, the detected periodicities extracted from the temperature anomalies, as well as their in-phase behavior with the polar motion, are more than compatible with what we have recently highlighted by comparing the same polar motion with the global volcanic eruption series (see \LeMouel \cite{LeMouel2023b}). Whether this is a coincidence will be the topic of further work.
  
\newpage

\appendix
\section{\label{sec:appendixA} \textbf{iSSA} components 3 to 6 (or 4 to 7 if Chandler counted but not common).}	
\counterwithin{figure}{section}
\setcounter{figure}{0}   
  
\begin{figure}[H]
    \centering
    \includegraphics[width=0.8\columnwidth]{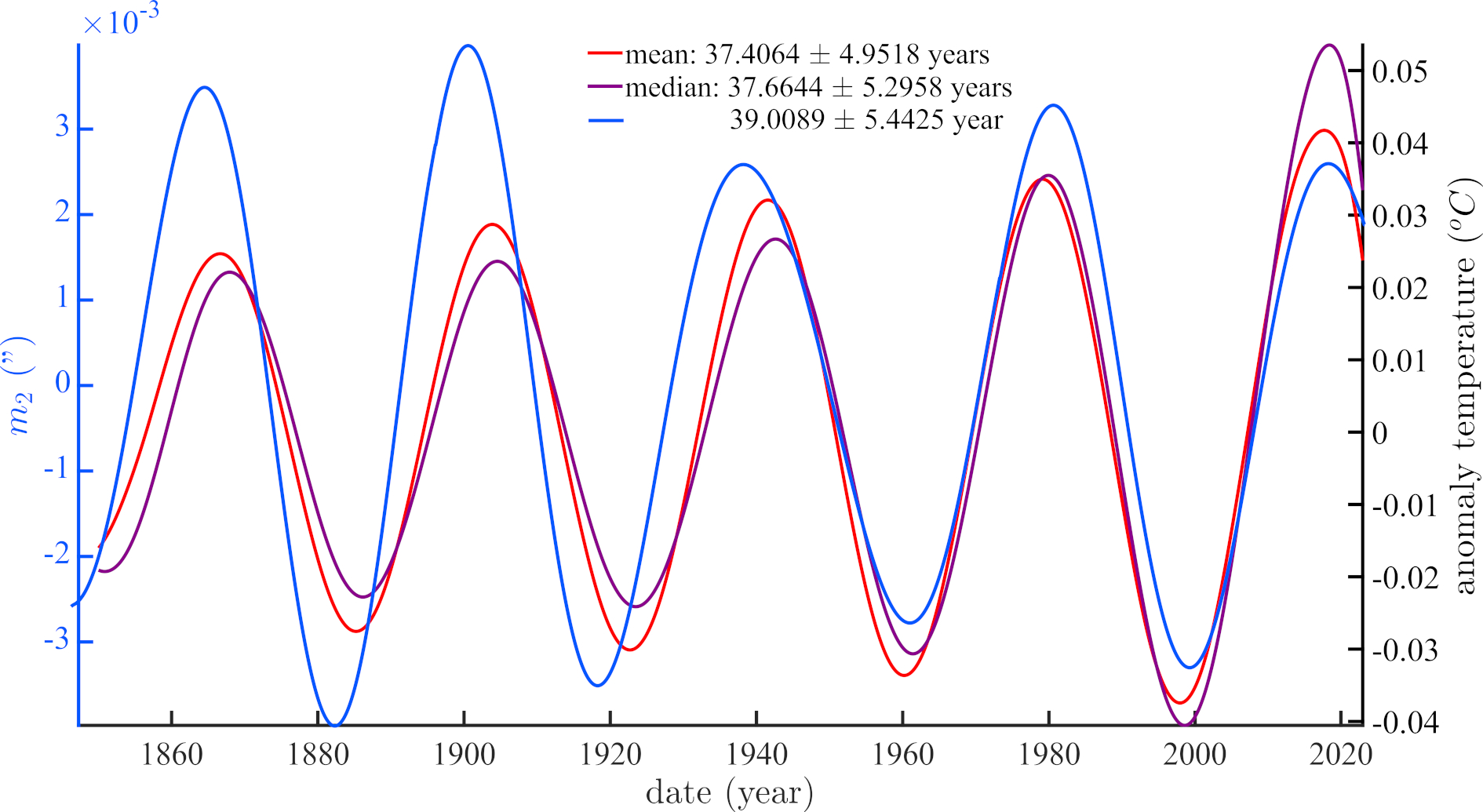}	
    \caption{Superposition et comparaison des composantes $\sim$40 ans extraites de  (courbe bleue) et des températures  médiane (courbe violette) et moyenne (courbe rouge) de la température du HadCRUT5.}
	\label{Fig:A01}
\end{figure}	

\begin{figure}[H]
    \centering
    \includegraphics[width=0.8\columnwidth]{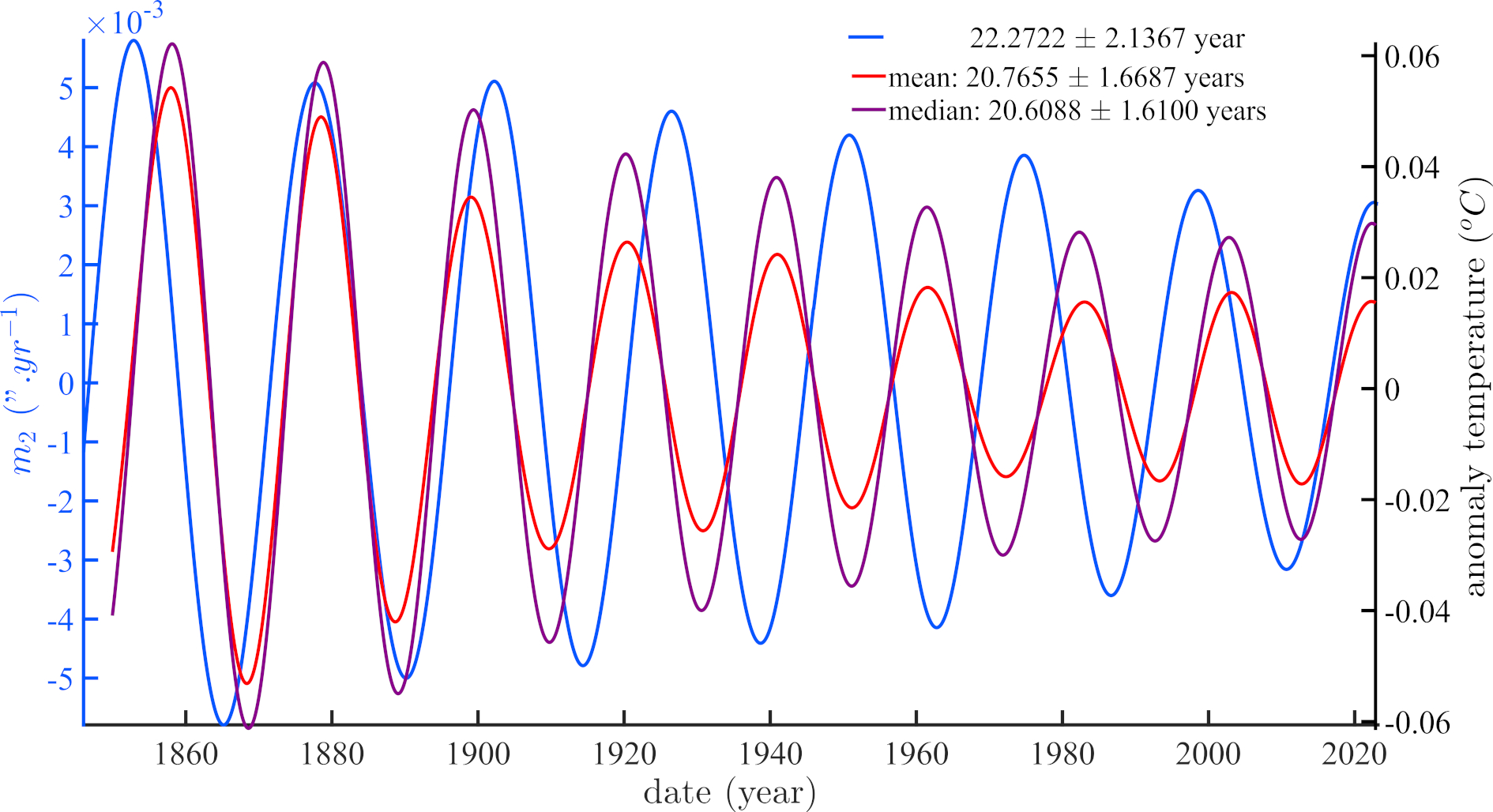}	
    \caption{Superposition et comparaison des composantes $\sim$22 ans extraites de  (courbe bleue) et des températures  médiane (courbe violette) et moyenne (courbe rouge) de la température du HadCRUT5.}
	\label{Fig:A02}
\end{figure}	
\newpage

\begin{figure}[H]
    \centering
    \includegraphics[width=0.8\columnwidth]{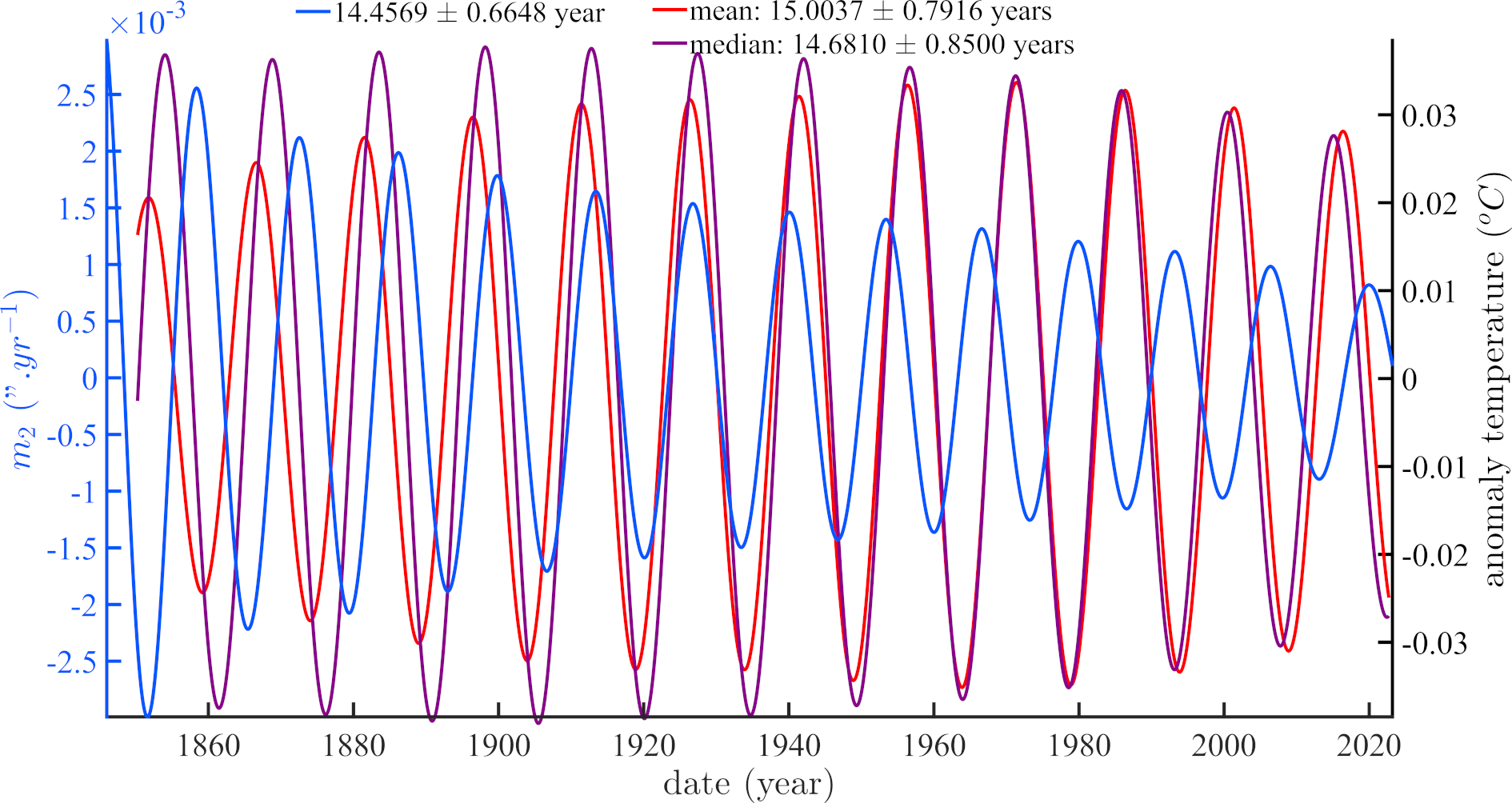}	
    \caption{Superposition et comparaison des composantes $\sim$15 ans extraites de  (courbe bleue) et des températures  médiane (courbe violette) et moyenne (courbe rouge) de la température du HadCRUT5.}
	\label{Fig:A03}
\end{figure}	

\begin{figure}[H]
    \centering
    \includegraphics[width=0.8\columnwidth]{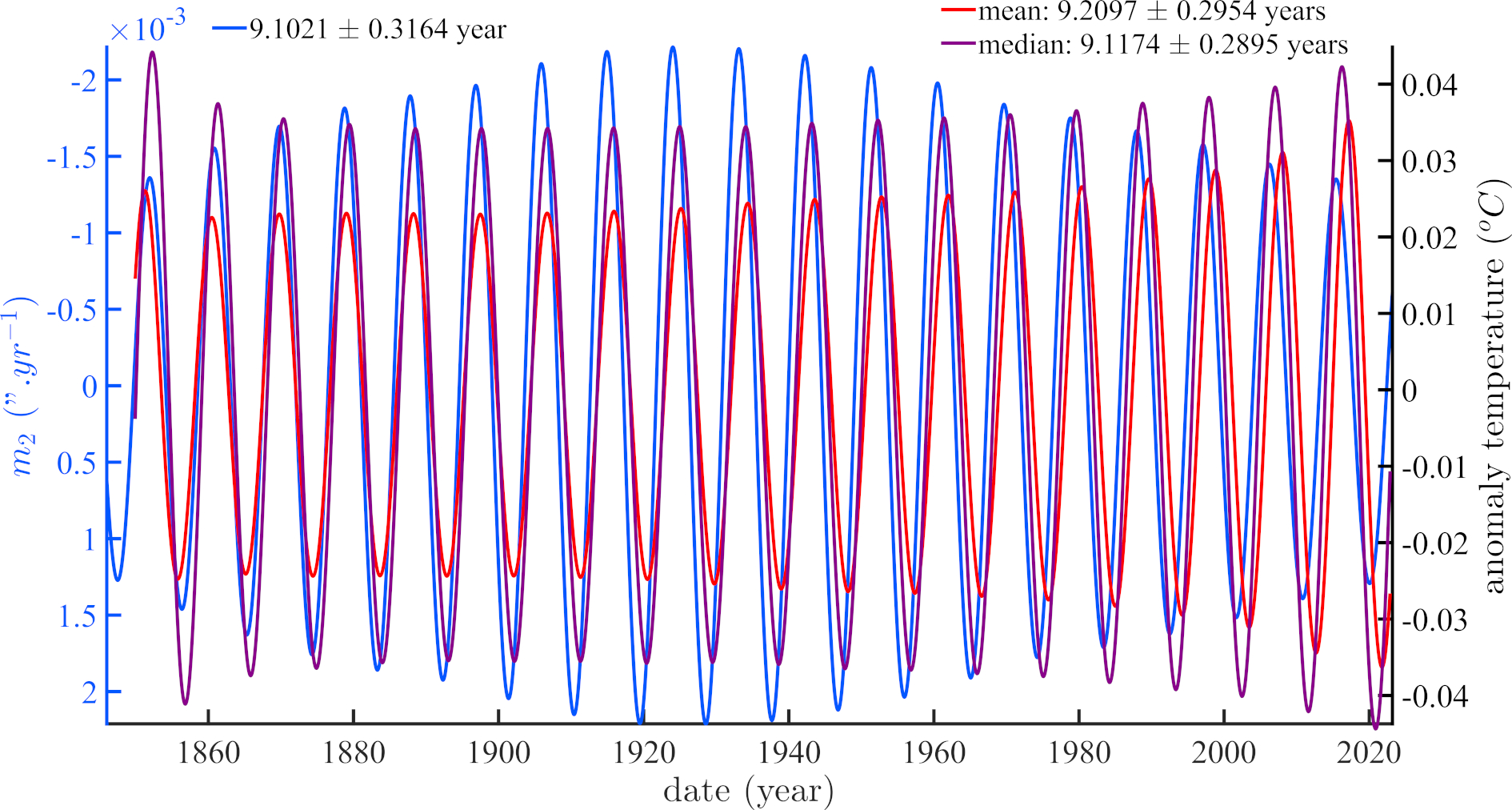}	
    \caption{Superposition et comparaison des composantes $\sim$9 ans extraites de  (courbe bleue) et des températures  médiane (courbe violette) et moyenne (courbe rouge) de la température du HadCRUT5.}
	\label{Fig:A04}
\end{figure}	  
  
\newpage
\bibliographystyle{ieeetr}
\bibliography{laplace_temp}
\end{document}